\documentclass[conference]{IEEEtran}
\IEEEoverridecommandlockouts
\usepackage{cite}
\usepackage{amsmath,amssymb,amsfonts}
\usepackage{algorithmic}
\usepackage{graphicx}
\usepackage{textcomp}
\usepackage{xcolor}
\def\BibTeX{{\rm B\kern-.05em{\sc i\kern-.025em b}\kern-.08em
    T\kern-.1667em\lower.7ex\hbox{E}\kern-.125emX}}
\begin{document}

\title{A Fast Graph Kernel Based Classification Method for Wireless Link Scheduling on Riemannian Manifold\\
}

\author{\IEEEauthorblockN{Rashed Shelim and Ahmed S. Ibrahim}
\IEEEauthorblockA{\textit{Department of Electrical and Computer Engineering} \\
\textit{Florida International University}\\
Miami, USA \\
rshel019@fiu.edu, aibrahim@fiu.edu}

}

\maketitle

\begin{abstract}
In this paper, we propose a novel graph kernel method for the wireless link scheduling problem in device-to-device (D2D) networks on Riemannian manifold. The link scheduling problem can be considered as a binary classification problem since each D2D pair can only hold the state active or inactive. Our goal is to learn a novel metric that facilitates the design of an efficient but less computationally demanding machine learning (ML) solution for the binary classification task of link scheduling problem that requires no channel state information (CSI) and a fewer number of training samples as opposed to other benchmark ML algorithms. To this aim, we first represent the wireless D2D network as a graph and model the features of each D2D pair, including its communication and interference links,  as regularized (i.e., positively-shifted) Laplacian matrices which are symmetric positive definite (SPD) one. By doing so, we represent the feature information of each D2D pair as a point on the SPD manifold, and we analyze the topology through Riemannian geometry. We compute the Riemannian metric, e.g., Log-Euclidean metric (LEM), which are suitable distance measures between the regularized Laplacian matrices. The LEM is then utilized to define a positive definite graph kernel for the binary classification of the link scheduling decisions. Simulation results demonstrate that the proposed graph Kernel-based method is computationally less demanding and achieves a sum rate of more than 95\% of benchmark algorithm FPLinQ \cite{b1} for 10 D2D pairs without using CSI and less than a hundred training network layouts.
\end{abstract}


\section{Introduction}
Wireless link scheduling, with full frequency reuse, in device-to-device (D2D) networks is one of the most fundamental problems in wireless communications. With the goal of the link scheduling to maximize the sum-rate by activating only a subset of mutually interfering D2D links at any given time, such a problem can be defined as a non-convex combinatorial optimization problem, which is an NP-hard one \cite{b2}. Traditional link scheduling approaches are usually based on non-convex optimization, e.g., Greedy heuristic search \cite{b3} or on the sequential link selection algorithm \cite{b4}. On the other hand, the state-of-art fractional programming approach (referred to as FPLinQ or FP) \cite{b1} iteratively solves the maximum sum-rate optimization problem within a finite number of iterations. However, these approaches require accurate CSI estimation, which is difficult for practical implementation for the densely deployed networks. 

As opposed to the requirement of instantaneous CSI, deep neural networks (DNN) have been recently employed for wireless link scheduling \cite{b2} which solely utilize the spatial location of D2D pairs as a proxy of CSI. However, the \textit{Spatial deep learning} method in \cite{b2} requires hundreds of thousands of training samples (i.e., network layouts), more precisely 800,000 samples in \cite{b2}, which 
requires a resource-intensive training process. Addressing this issue, a graph embedding-based DNN method is proposed in \cite{b5} which reduces the number of training layouts to 500 while maintaining the advantage of requiring no instantaneous CSI. In the graph embedding approach in \cite{b5}, each D2D pair is considered a single node and modeled as a vertex in a graph, whereas the interference link between every two D2D pairs is modeled as an edge. 
Recently, the underlying non-Euclidean structures of the wireless networks are studied to extract the unexplored interference characteristics of the D2D networks in \cite{b6}. In this work, a novel Riemannian metric, e.g., Log-Euclidean Metric (LEM), is used to characterize interference among the D2D pairs, and a sequential link selection algorithm is proposed for link scheduling. Together the innovative graph modeling of interference network in \cite{b5}, and the novel mapping of wireless link scheduling to non-Euclidean manifolds in \cite{b6} can lead to design more efficient but less computationally expensive Machine learning-based link scheduling solution that reduces the requirement of the number of training layouts even further, and this is the main motivation of this paper.

With this aim, we propose a general classification framework for wireless link scheduling decisions, which combines the geometric properties of the Riemannian manifold $\mathcal{M}$ via graph embedding of the D2D interference network to define graph kernel for binary classification of D2D pair state. The graph kernel is a kernel function that computes the inner product on the graph features of D2D pairs which can be intuitively understood as a function measuring the similarity between the two D2D pairs graph features. 
To this end, we first model the D2D interference network as a fully connected graph via graph embedding technique where the transmitter and receiver of each D2D pair is modeled as the vertex in the graph, and the corresponding communication and interference links are modeled as the edges. Then from this graph, we model the connectivity pattern of each D2D node with three separate Laplacian matrices 
to embed the features: 1) of the D2D communication link, 2) of interference links towards itself, and 3) of the interference it introduces to its neighbor D2D pairs. 
We represent the embedded feature information of each D2D pairs as a single point on the interior of the convex cone, or simply on a conic manifold \cite{b7}, which follows non-Euclidean geometry \cite{b8}. 

Conic or SPD structures are the special class of Riemannian manifolds \cite{b9} which are characterized by Riemannian metrics and studied by Riemannian geometry \cite{b10}. Riemannian metrics such as Log-Euclidean metric (LEM) \cite{b11} can be used as suitable measures of distance between the regularized Laplacian matrices of D2D pairs on the manifold. So, we intend to learn the LEM through
a simple (i.e., less computationally complex) machine learning (ML) solution for binary classification. However, Riemannian manifolds are non-linear, and the machine learning solutions available in the literature are in practice designed for $\mathbb{R}^n$, and thus cannot be directly applied on Riemannian manifold. 
However, many ML algorithms designed on $\mathbb{R}^n$ can be generalized to Hilbert space where the vector norms and inner products are defined \cite{b12}. 

We use the LEM to define a positive definite graph kernel which maps 
each D2D pair graph features to a feature vector in a higher dimensional Reproducing Kernel Hilbert Space (RKHS) to perform the inner products between the mapped feature vectors. This allows us to directly use the positive definite graph kernel with a simple machine learning solution such as a support vector machine (SVM) for the binary classification of wireless link scheduling, and this is the main contribution of this paper. The novelty of this contribution lies in the fact that we \textit{learn} the LEM by graph kernel, which yields a much richer representation of the original data distribution, and hence, helps to learn the underlying interference characteristics of the D2D network faster to classify the link scheduling decision. We demonstrate by simulation that our proposed graph Kernel-based method is less computationally demanding and achieves a sum rate of more than 95\% of the benchmark algorithm FPLinQ [1] (for 10 D2D pairs) by using only less than a hundred training network layouts without requiring any instantaneous channel state information.

The rest of this paper is organized as follows. The representation of graph features of D2D pairs as SPD matrices and the application of positive definite graph kernel on Riemannian manifold is presented in Section~\ref{system_model}. Section~\ref{problem_statement} discusses the problem formulation of maximizing the sum rate of D2D wireless communication networks over Riemannian manifold. Section~\ref{solution} presents the proposed graph kernel-based binary classification method for wireless link scheduling decisions. The simulation results are presented in Section~\ref{simulation}. Finally, the conclusion is provided in Section~\ref{conclusion}.

\section{System Model}\label{system_model}
In this section, we briefly introduce the notions of Riemannian geometry on the SPD manifolds, including its metics. Next, we present the system model starting with representing the feature information of D2D pairs as  point (i.e., SPD) on non-euclidean Riemannian manifold, and then discuss the use of graph kernel methods on the non-Euclidean SPD manifolds.
\subsection{Riemannian Geometry}
A differential manifold $\mathcal{M}$ is a topological one \cite{b13} that is locally similar to the $n$ dimensional Euclidean space $\mathbb{R}^n$. The tangent space $T_p\mathcal{M}$ 
at any point $p$ on the differential manifold $\mathcal{M}$ is a vector space of all possible tangent vectors passing through the point $p$. The Riemannian manifold $(\mathcal{M},d)$ is a differential manifold $\mathcal{M}$ with metric $d$ which is studied by Riemannian geometry \cite{b9},\cite{b10}. The $n \times n$ SPD matrices lie on the interior of conic manifold \cite{b7}, which is a special class of Riemannian manifold \cite{b9}. Let, $\textit{Sym}_n^{++}$ denote the set of all symmetric positive definite matrices of size $n \times n$. The smooth inner product of all tangent space is known as the Riemannian metric, and it captures the geometric properties on the manifold, such as the geodesic distance (i.e., the shortest curves) between the two points on the manifold. The two most popular distance measures over Riemannian manifold are the Affine-Invariant Metric (AIM) \cite{b14}, and Log-Euclidean Metric (LEM) \cite{b11} which truly measure the geodesic distances.      
\subsection{Modeling D2D pairs features over Riemannian Manifold}
\begin{figure}[htbp]
\centerline{
\includegraphics[width=4cm, height= 5 cm]{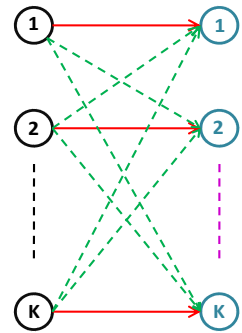}}
\caption{\small{K-user interference channel.}}
\label{K_user_interference}
\end{figure}

Fig.~\ref{K_user_interference} shows a fundamental $K$-user interference channel model which is the underlying model for the wireless networks of $K$ D2D pairs. Each D2D pair $D_q$, $q \in K$ consists of a transmitter (marked by black) and a receiver (marked by blue). The communication links are illustrated with solid red lines and the interference links are illustrated with dashed green line. 
With full frequency reuse, the communication between any D2D pair $D_q$ causes interference to the receivers of its neighbor D2D pairs $D_i$, $i \in K$ and $i \neq q$. The model can be represented as a weighted and directed finite graph $G(V,E)$, 
with $n = 2K$ nodes and $m$ edges. For an edge $l$ connecting nodes $i$ and $j$, where $i,j \in K$, we define the edge vector $\boldsymbol{a}_l \in \mathbb{R}^n$ as $a_{l_i}= 1$, $a_{l_j}= -1$ and rest of the entries are zero. The $l$-th column of the incidence matrix $\boldsymbol{A} \in \mathbb{R}^{n \times m}$ of graph $G$ is given by the edge vector $\boldsymbol{a}_l$. On the other hand, the weight matrix $\boldsymbol{W} \in \mathbb{R}^{m \times m}$ is a diagonal matrix where the diagonal elements are derived by the weights of the $l$-th edge. Noting that for the wireless network, the channel gains are a function of the distance induced path-loss and the geographical location information of D2D pairs is sufficient as a proxy of complete knowledge on CSI \cite{b2}, the weight matrices of any D2D pairs can be formed by the Euclidean distances between each D2D pairs and corresponding to its neighbor D2D nodes. Using the incidence and weight matrix, the Laplacian matrix $\boldsymbol{L}\in \mathbb{R}^{n \times n}$ is computed as $\boldsymbol{L} = \boldsymbol{A}\boldsymbol{W}\boldsymbol{A}^T,$ where, $T$ denotes the matrix transposition. The Laplacian matrices are positive semi-definite. With a simple regularization step \cite{bspd} by adding  a scaled identity matrix results in a regularized SPD Laplacian matrix as $\boldsymbol{S} = \boldsymbol{A}\boldsymbol{W}\boldsymbol{A}^T + \gamma \boldsymbol{I}$, where $\boldsymbol{I}$ is the $n \times n$ identity matrix $\gamma > 0$ is a regularization parameters. 


\subsection{Graph Kernel Method on the Riemannian Manifold of SPD Matrices}
In practice, the input data are not often separated enough for classification due to the non-linearity of the decision boundary. Kernel methods overcome this issue by mapping the input data into a high-dimensional feature space where the classification task is performed. Noting that many classification algorithms (e.g. Support Vector Machine) originally designed on Euclidean space $\mathbb{R}^n$ can be directly generalized to Hilbert spaces where the vector norms and inner products are defined\cite{b12}, the kernel method can be generalized to Riemannian manifold as shown in Fig.~\ref{Hilbert_space}. The concept is as follows: 
\begin{figure}[htbp]
\centerline{
\includegraphics[width=8cm, height= 5 cm]{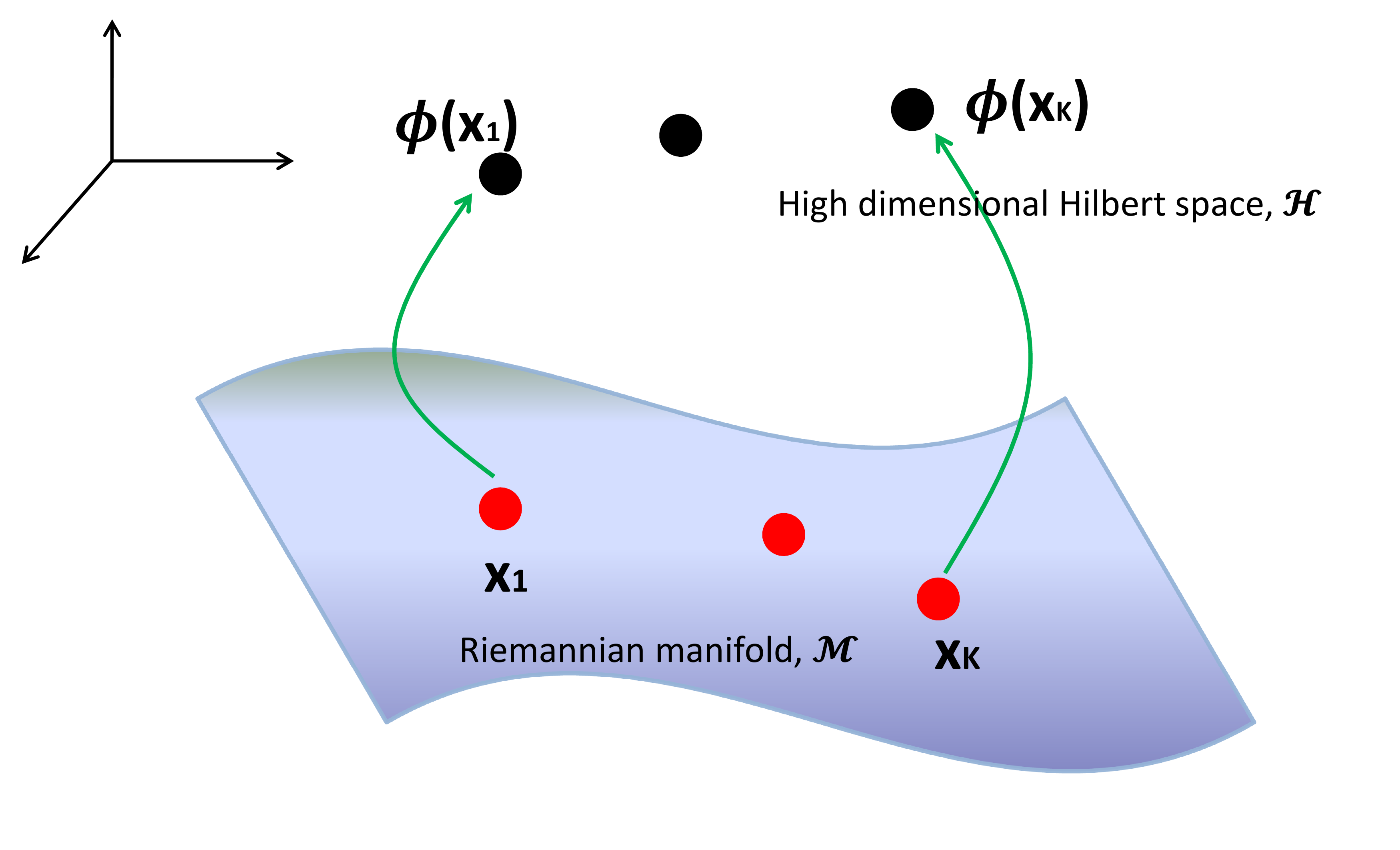}}
\caption{\small{Mapping the points (i.e., $S_{D_q}$) on Riemannian manifold $\mathcal{M}$ to the feature vectors (i.e., $\phi(x)$) in a Hilbert space $\mathcal{H}$.}}
\label{Hilbert_space}
\end{figure}
Each point $x$ on the Riemannian manifold $\mathcal{M}$ that we represent by regularized Laplacian matrix $\boldsymbol{S}_{D_q}$ is mapped to a feature vector $\phi(x)$ in a Hilbert space $\mathcal{H}$. A kernel function $\mathcal{F}_K : (\mathcal{M} \times \mathcal{M}) \rightarrow \mathbb{R}$ is used to perform the inner product between the points (i.e., correlation between the SPDs) on manifold on the space $\mathcal{H}$ thus functioning it as a Reproducing Kernel Hilbert Space (RKHS) \cite{b12}. However, according to Mercer's theorem, only the positive definite kernels can define the valid RKHS. So, for our binary classification task of link scheduling, we exploit the LEM \cite{b11}, a.k.a., log-Frobenius distance, as a Riemannian metric to define a positive definite kernel \cite{b12}. The LEM distance between $\boldsymbol{S}_{D_q}$ and $\boldsymbol{S}_{D_{q'}}$, where $q,q' \in K$, and $q \neq q'$ can be computed as 
\begin{equation}
\mathcal{L}(\boldsymbol{S}_{D_q},\boldsymbol{S}_{D_{q'}}) = \| \log(\boldsymbol{S}_{D_q}) - \log(\boldsymbol{S}_{D_{q'}}) \|_F^2 \label{LEM}
\end{equation}
where, $\|.\|_F$ denotes the Frobenious matrix norm.

Therefore, the LEM distance in \eqref{LEM} can be learned as a Riemannian metric for the binary classification problem of link scheduling, as will be discussed in the next two sections.

\section{Problem Formulation}\label{problem_statement}
In this section, we formulate the fundamental problem of sum-rate maximization of the K-user interference channel, as shown in Fig.~\ref{K_user_interference}, and later map it to the formulation over SPD manifolds. 
Let, $d_q \in \{0,1\}$ be a variable of binary decision which represent the state of $q$-th D2D pair (i.e., active or inactive), $q = 1,\ldots,K$, with $d_q = 1$ if $D_q$ is activated and the communication link is scheduled, and vice-versa. Let, $\boldsymbol{d}= {[d_1, \ldots, d_k]}^{\boldsymbol{T}}$ be a $K \times 1$ vector that contains all these variables of binary link scheduling decisions. Activating all the D2D links simultaneously will result in a poor data rate due to the interference between the links. Thus, we aim to find the optimal combinations of such binary decisions that maximize the summation of the individual information-theoretic rates over bandwidth $B$ as given by
\begin{align}
    \max_{\boldsymbol{d}} &\sum_{q=1}^K B \log_2 \bigg( 1 + \frac{pd_q{|h_{qq}|}^2 \rho_{qq}^{-\alpha}}{\sum_{i\neq q}pd_i{|h_{iq}|}^2 \rho_{iq}^{-\alpha}+\sigma^2} \bigg),\nonumber\\
    &\text{s.t.}\text{ } d_q \in \{0,1\}, q = 1,2,\ldots,K,\label{problem_formulation}
\end{align}
where $p$ is the transmission power of which we assume to be the same for all links, $h_{iq}$ and $\rho_{iq}$ represents the fast-fading channel gain and the Euclidean distance, respectively, between the $i$-th transmitter and $q$-th receiver. Moreover, $\alpha$ denotes the path loss exponent, and $\sigma^2$ denotes the noise variance. 

The optimization problem in \eqref{problem_formulation} is a challenging non-convex combinatorial problem, which is generally NP-hard since the optimal combination of the values of binary decision variables $d_q$ depends on the choices of other ones \cite{b1}. By representing features of each D2D pairs as a point on Riemannian manifold through the regularized Laplacian matrices $S_{D_q}$, $q = 1, \ldots, K$, we aim to find the optimal values of each binary variables by learning the Riemannian metric LEM through our proposed positive definite graph kernel $\mathcal{F}_K$ based ML solution to perform the inner-products between the SPDs by mapping it to Hilbert space and thus separate the features vectors of each D2D pair $D_q$ as active and inactive classes, as in
\begin{equation}
    \mathcal{F}_K: (\textit{Sym}_n^{++} \times \textit{Sym}_n^{++})  \rightarrow d_q, 
\end{equation}
where $q = 1, 2, \ldots, K$ and $d_q \in \{0,1\}$.

\section{Wireless Link Scheduling by Graph Kernel based Classification Method on Riemannian Manifold}\label{solution}
\begin{figure}[htbp]
\centerline{
\includegraphics[width=9cm, height= 5 cm]{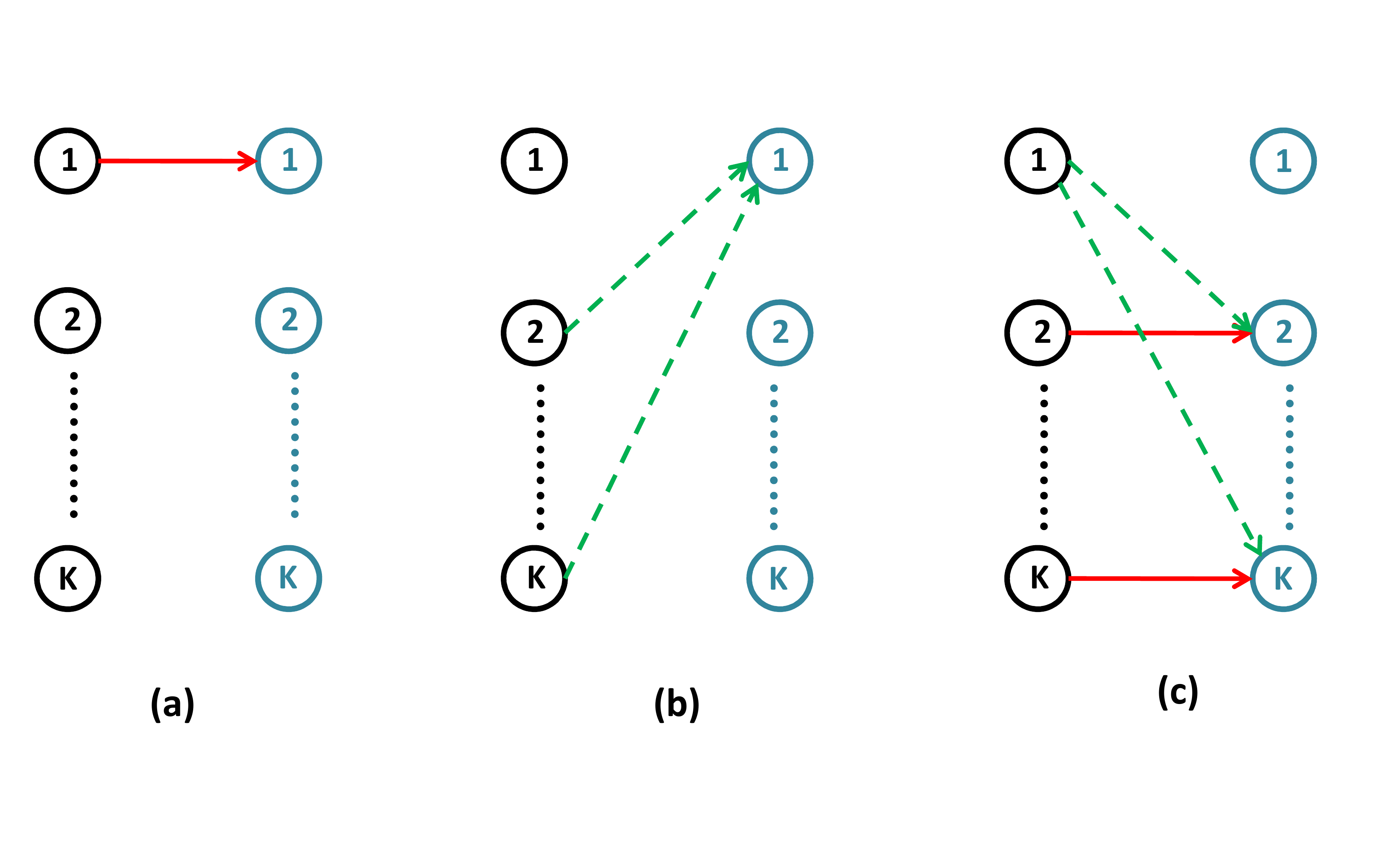}}
\caption{\small{Example of graph features embedding of $D_{q=1}$ D2D pair: a) embedding the features of desired link, b) embedding the features of interference links towards itself, and c) embedding the feature information of interfering links towards its neighbor.}}
\label{Three_graph}
\end{figure}

In this section, we describe the graph kernel based binary classification method for wireless link scheduling. 
Using the graph modeling technique described in previous section, we embed the feature information of each D2D pair, $D_q$, $q \in K$ on the manifold as follows: we first consider a special case of graph $G$ is the one with a single direct-link (i.e., communication link), denoted as $G_\text{com} (V, E_\text{com})$ for the $q$-th D2D pair, where $q = 1, \ldots, K$ and $E_q$ includes only singe edge from $q$-th transmitter to $q$-th receiver as shown in Fig.~\ref{Three_graph}a. The incidence matrix of this case $\boldsymbol{A}_\text{com} \in \mathbb{R}^{n \times 1}$ has 1 and -1 at its $q$-th transmitter and receiver positions, respectively, and the rest of the entries are zero. The weight matrix $\boldsymbol{W}_\text{com} \in \mathbb{R}^{1 \times 1}$ of this special case has a single entry which is the Euclidean distance between the D2D pair. Hence, the Laplacian matrix $ \boldsymbol{L}_\text{com}$ for the communication link feature $D_q$ is given by $\boldsymbol{L}_\text{com} = \boldsymbol{A}_\text{com}\boldsymbol{W}_\text{com}\boldsymbol{A}_\text{com}^T.$

Next, we consider a second special case of graph $G$ is the one with the interference links from all the transmitters of neighbor D2D pairs $D_i$, $i \in K$ and $i \neq q$ to the receiver of the D2D pair $D_q$, $q \in K$ (Fig.~\ref{Three_graph}b), denoted as $G_\text{int}(V, E_\text{int})$. It characterizes the interference the D2D pair $D_q$ receives from it's neighbor D2D pairs $D_i$ if the neighbor pairs the scheduled. Here $E_\text{int}$ includes the interference links from the all neighbor transmitters to the receiver of $D_q$. The incidence matrix of this case $\sum_i\boldsymbol{A}_\text{int} \in \mathbb{R}^{n \times (K-1)}$ has 1 and -1 at it's $i$-th transmitters, where $i \in K$ and $i \neq q$, and $q$-th receiver positions, respectively, and rest of the entries are zero. The diagonal weight matrix $\boldsymbol{W}_\text{int} \in \mathbb{R}^{(K-1) \times (K-1)}$ of this case has diagonal entries which are the Euclidean distances between the transmitters of $D_i$ to the reciver of $D_q$. Thus, the Laplacian matrix $ \boldsymbol{L}_\text{int}$ for the interference link feature of $D_q$ can be expressed as $ \boldsymbol{L}_\text{int} = \sum_i\boldsymbol{A}_\text{int}\boldsymbol{W}_\text{int}\big({\sum_i\boldsymbol{A}_\text{int}}\big)^T.$

Finally, we consider a third special case of graph $G$ to embed the features of neighbor D2D pairs (Fig.~\ref{Three_graph}c), denoted as $G_\text{nbr}(V, E_\text{nbr})$, where $E_\text{nbr}$ includes all the direct-links (i.e., communication links) of the neighbor D2D pairs $D_i$, $i \in K$ and $i \neq q$ and the interference links from the the transmitter of the D2D pair $D_q$, $q \in K$ to the receivers of all it's neighbor $D_i$. Basically, it captures the impacts of the interference to it's neighbors if the communication link of $D_q$ is scheduled. The incidence matrix of this case $\sum_i\boldsymbol{A}_\text{nbr} \in \mathbb{R}^{n \times (n-2)}$, where $i \in K$ and $i \neq q$, has 1 and -1 at it's $i$-th transmitter-receiver pair positions along with the 1 and -1 at it's $q$-th transmitter and $i$-th receivers positions, respectively. Rest of the entries are zero. The diagonal weight matrix $\boldsymbol{W}_\text{nbr} \in \mathbb{R}^{(n-2) \times (n-2)}$ of this case has diagonal entries which are the Euclidean distances between the nodes of $D_i$ pairs, and between the $D_q$ pair's transmitter to the receiver of all neighbor D2D pairs $D_i$. Thus, the Laplacian matrix $ \boldsymbol{L}_\text{nbr}$ can be expressed as $\boldsymbol{L}_\text{nbr} = \sum_i\boldsymbol{A}_\text{nbr}\boldsymbol{W}_\text{nbr}\big({\sum_i\boldsymbol{A}_\text{nbr}}\big)^T.$

The Laplacian matrices $\boldsymbol{L}_\text{com}$, $\boldsymbol{L}_\text{int}$ and $\boldsymbol{L}_\text{nbr}$ are guaranteed to be always positive semi-definite and can be regularized to make it symmetric positive definite (SPD) \cite{bspd} by 
\begin{align}
\boldsymbol{S}_\text{com} &= \boldsymbol{A}_\text{com}\boldsymbol{W}_\text{com}\boldsymbol{A}_\text{com}^T + \gamma \boldsymbol{I},\\
 \boldsymbol{S}_\text{int} &= \sum_i\boldsymbol{A}_\text{int}\boldsymbol{W}_\text{int}\big({\sum_i\boldsymbol{A}_\text{int}}\big)^T +\gamma \boldsymbol{I}, \\
 \boldsymbol{S}_\text{nbr} &= \sum_i\boldsymbol{A}_\text{nbr}\boldsymbol{W}_\text{nbr}\big({\sum_i\boldsymbol{A}_\text{nbr}}\big)^T +\gamma \boldsymbol{I},
\end{align}
where, $i \in K$, $i \neq q$, and $\gamma > 0$ is a regularization parameters and $\boldsymbol{I}$ is the $n \times n$ identity matrix. Addressing that the sum of symmetric positive definite matrices is also a positive definite matrix \cite{b6_1}, we add these three regularized Laplacian matrices to embed the complete feature information of each D2D pairs $D_q$ and thus form a single regularized Laplacian Matrix which is given by
\begin{equation}
    \boldsymbol{S}_{D_q} = \boldsymbol{S}_\text{com} + \boldsymbol{S}_\text{int} + \boldsymbol{S}_\text{nbr}.
\end{equation}
Hence, we represent the embedded feature information of each D2D pair as a single point on the interior of the convex cone \cite{b7}, which are the special class of Riemannian manifold \cite{b8}.

Now, given a set of D2D pair feature graphs modeling on Riemannian manifold $\{S_{D_q}^l \in \textit{Sym}_n^{++}, q = 1, 2, \ldots, K , l = 1,2,\ldots,T\}$, where $K$ is the total number of D2D pairs in a wireless network layout $l$ and $T$ denotes the total number of wireless network layout samples, our aim is to classify the D2D feature graph representations on manifold between active and inactive state for link scheduling. With $S_{D_q}^l \in \textit{Sym}_n^{++}$ for all $l$ and utilizing the Riemannian Metric LEM $\mathcal{L}(\boldsymbol{S}_{D_q}^l,\boldsymbol{S}_{D_{q'}}^l)$, the positive definite graph kernel can be defined as \cite{b12}
\begin{equation}
 \mathcal{F}_K (\boldsymbol{S}_{D_q}^l,\boldsymbol{S}_{D_{q'}}^l) = \exp \Bigg(- \frac{\mathcal{L}^2(\boldsymbol{S}_{D_q}^l,\boldsymbol{S}_{D_{q'}}^l)}{\gamma^2} \Bigg), \forall \gamma \neq 0.
\end{equation}

Having defined the positive definite graph kernel, we incorporate the kernel Support Vector Machine (SVM) for our binary classification task. We apply a supervised training approach to optimize the parameters, and we utilize the FPLinQ \cite{b1} solutions as training targets (i.e., D2D links states) to classify the link scheduling decision. The classification process can be described as follows: For a given set of \textit{training} examples $\{ (\boldsymbol{S}_{D_q}^l, t_q^l) \}_1^K$, where $t_q^l \in \{0,1\}$, $q = 1, \ldots, K$ represents the target (i.e., label) of $q$-th D2D pair's link belonging to $l$-th training D2D wireless network layout sample, the graph kernel-based SVM searches for a hyperplane in Hilbert space $\mathcal{H}$ that separates the feature vectors (mapped from the D2D  feature graphs representation on the manifold) to the active and inactive classes (i.e., 1 as active class and 0 as inactive class) with maximum margin. On the other hand, the class of a test point $\boldsymbol{S}_{D_q}^{l_t}$ of $q$-th link belonging to $l_t$-th test D2D wireless network layout sample is determined by the position of the feature vector $\phi (\boldsymbol{S}_{D_q}^{l_t})$ in the Hilber space $\mathcal{H}$ relative to the separating hyperplane. Hence, the optimal combination of scheduling decisions $d_q^{l_t} \in \{0,1\}$, $q = 1, \ldots, K$ of test D2D wireless network layout $l_t$ are obtained by our proposed graph kernel-based binary classification method.

\section{Simulation Results} \label{simulation}
This section presents the simulation results of our proposed graph kernel-based link scheduling method and compares it against other benchmark solutions. We deploy $K=10$ D2D pairs on the square-shaped network layout, where we use different field lengths of the layout such as 350m, 400m, 450m, and 500m. The 10 transmitters are deployed by the uniform distribution. At the same time, the 10 receivers are placed within the disk around its corresponding transmitter pair with a uniformly distributed radius between 2m to 65m. In our simulation, we randomly generate the fast-fading channel, and we consider a distance-based path-loss according to the outdoor model ITU-1411 without any shadowing effect. Rest of simulation parameters are kept same as in \cite{b1},\cite{b2}, \cite{b5} and \cite{b6} and included in Table \ref{tab:parameter} to have a fair comparison.  
\begin{table}[htbp]
\centering
\caption{Network Simulation Parameters.}
\begin{tabular}{|c|c|}
\hline
\textbf{Parameter} & \textbf{Value}\\
\hline
D2D distance $(d_{\min}, d_{\max})$ & 2m to 65m\\
\hline
Carrier Frequency & 2.4 GHz\\
\hline
Bandwidth, B & 5MHz\\
\hline
Transmit power $p$ of the activated link & 40 dBm\\
\hline
Antenna height (both Transmitter and Receiver) & 1.5m\\
\hline
Antenna gain (both Transmitter and Receiver) & 2.5 dB\\
\hline
Noise spectral density & -169 dBm/Hz\\
\hline
\end{tabular}
\label{tab:parameter}
\end{table}

Table \ref{tab:compare} enlisted different simulated schemes that we use for comparison. The FPlinQ \cite{b1} scheme iteratively solves the maximum sum-rate optimization problem within a finite number of iterations by decoupling the signal and interference terms in \eqref{problem_formulation}. The greedy algorithm employs a sequential link scheduling algorithm where it only adds a new link if the addition does not reduce the sum rates of the already scheduled D2D links. The strongest link scheme only activates the links with maximum SNR. The FPlinQ, greedy and strongest link schemes require CSI for scheduling. The LEM scheme \cite{b6} measures the interference among the D2D pairs by LEM, and a sequential link selection algorithm is proposed for link scheduling on the Riemannian manifold. The random algorithm randomly activates the links, while the all active scheme activates all D2D links. The last three algorithms mentioned above do not require any CSI. 
\begin{table}[htbp]
\centering
\caption{Simulated benchmark link scheduling schemes.}
\begin{tabular}{|c|c|}
\hline
\textbf{Scheme} & \textbf{CSI}\\
\hline
GKernel method(proposed) & No\\
\hline
LEM \cite{b6} & No\\
\hline
FPLinkQ \cite{b1} & Yes\\
\hline
Greedy & Yes\\
\hline
Strongest Link Only & Yes\\
\hline
Random & No\\
\hline
All Active & No\\
\hline
\end{tabular}
\label{tab:compare}
\end{table}

\begin{figure}[htbp]
\centerline{
\includegraphics[width=9cm, height= 6cm]{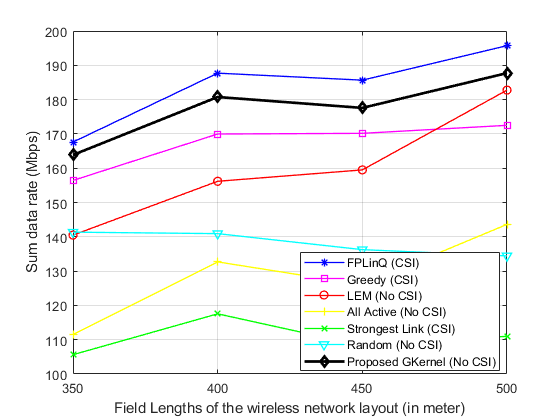}}
\caption{\small{Achievable sum rate by various wireless D2D link scheduling algorithm for $K=10$ links.}}
\label{Sum_rate}
\end{figure}

Fig.~\ref{Sum_rate} compares the achievable sum rate by different D2D link scheduling schemes under a fast-fading scenario. As shown in the figure, both FPlinQ and greedy algorithms utilize the CSI information to achieve a high sum rate. In contrast, while our proposed graph kernel method does not utilize CSI, it achieves a sum rate of more than 95\% compared to FPlinQ (for K = 10 links at field length of 500m).  It is also shown that the proposed graph kernel method outperforms both the greedy and LEM-based sequential algorithms \cite{b6} and achieves a significantly higher sum rate than the strongest link and all other algorithms that do not require CSI.

\begin{figure}[htbp]
\centerline{
\includegraphics[width=4.5cm, height= 4cm]{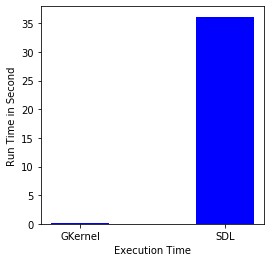}}
\caption{\small{Execution time of testing with 10 layouts for K=10 links at layout field length of 500m.}}
\label{execution_time}
\end{figure}

Table \ref{tab:MLcompare} compares performance of various machine learning solutions for link scheduling for $K=10$ links at 500m $\times$ 500m wireless network layouts. Though the SDL \cite{b2} and graph embedding \cite{b5} schemes do not require CSI, they need 800,000 and 500 \textit{training layouts}, respectively, to achieve the sum rate above 95\% of FPLinQ \cite{b1}. In contrast, our proposed graph kernel-based method only requires 90 training layouts to achieves a similar performance while requiring no CSI. We also compare the execution time of testing our proposed scheduling method against the SDL with 10 layouts for $K=10$ links and layout field length of 500m in the same simulation platform. As shown in Fig.~\ref{execution_time}, our proposed method takes only 0.091 seconds as opposed to the SDL schemes, which takes 36.147 seconds to perform the testing. The results can be linked to the computational complexity, which shows that our proposed method is computationally less demanding. Both of these results indicate that our proposed graph kernel-based link scheduling algorithm is \textbf{faster} to learn the features of D2D pairs, which requires only less than a hundred wireless network layouts for training and significantly less amount time for testing. This is one of the main contributions of this paper.
\begin{table}[htbp]
\centering
\caption{Average sum rate as \% of FPlinQ of various ML schemes for K=10 links (at field length 500 meter).}
\begin{tabular}{|c|c|}
\hline
\textbf{Scheme} & \textbf{Sum rate} \\
\hline
SDL \cite{b2} & 95.50\\
\hline
Graph Embedding \cite{b5} & 97.51\\
\hline
GKernel (proposed)  & 95.91\\
\hline
\end{tabular}
\label{tab:MLcompare}
\end{table}

\begin{figure}[htbp]
\centerline{
\includegraphics[width=9cm, height= 6cm]{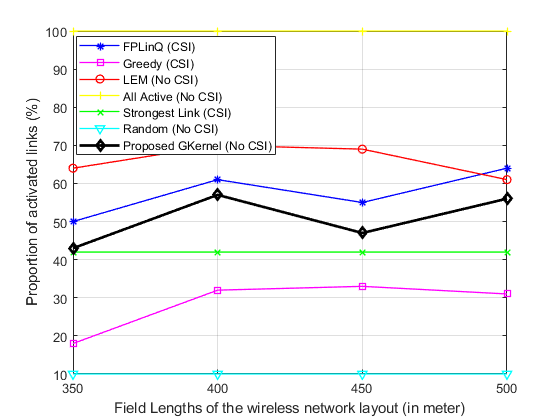}}
\caption{\small{Proportion of activated links (scheduling) of D2D pairs for various wireless link scheduling algorithms.}}
\label{activation}
\end{figure}

Finally, to understand the behavior of the activation pattern of the simulated link scheduling scheme for different field lengths, Fig.~\ref{activation} illustrates the proportion of activated links. As shown, the proposed LEM-based graph kernel method closely follows the patterns of link activation ratio of FPlinQ while not utilizing any CSI.

\section{Conclusion}\label{conclusion}
In this paper, we propose a mathematical framework based on Riemannian geomatry and a novel graph kernel based SVM method for wireless link scheduling problem in D2D communications. First, we represent the graph feature information of each D2D pair as points on a conic manifold, a special class of Riemannian manifold, through the regularized Laplacian martrix which is a symmetric positive define ones. Then, we learn the novel Riemannian metic LEM though our proposed positive definite graph kernel for the binary classification of wireless link scheduling decision. To best of our knowledge, this paper is the first attempt to introduce graph the kernel based classification method on Riemannian manifold for wireless D2D link scheduling problem. We show that our proposed graph Kernel based SVM method is faster in learning the graph features of D2D pairs through LEM and achieves a sum rate of more than 95\% of benchmark algorithm FPLinQ \cite{b1} for 10 D2D pairs by only utilizing the spatial locations of D2D pairs and less than a hundred training network layouts.    

\vspace{12pt}
\end{document}